\documentstyle[11pt,newpasp,twoside,epsf]{article}
\def\edcomment#1{\iffalse\marginpar{\raggedright\sl#1\/}\else\relax\fi}
\reversemarginpar

\begin{document}
\title{Properties of the circumnuclear medium in Active Galactic Nuclei}
 \author{R. Gilli, G. Risaliti, P. Severgnini}
\affil{Dipartimento di Astronomia e Scienza dello Spazio, Universit\`a
di Firenze, L.~E.~Fermi~5, 50125 Firenze, Italy}
\author{R. Maiolino, A. Marconi, M. Salvati}
\affil{Osservatorio Astrofisico di Arcetri, L.~E.~Fermi~5, 50125 Firenze,
Italy}

\begin{abstract}
We propose a classification scheme for the circumnuclear medium
of AGNs based on their location in the $A_{\rm V}-N_{\rm H}$ 
plane (related to dust absorption and gas absorption, respectively).
We present evidence that the $A_{\rm V}/N_{\rm H}$ ratio is often
lower than the Galactic value. This is shown by the broad
line ratios of a sample of intermediate Seyferts, and by the optical/X--ray
ratio of line selected (as opposed to color selected) AGNs. We point out
the implications for the physical properties of the absorber. A complete
classification would require a third axis, related to the density of the
absorber and/or its distance from the AGN. 
\end{abstract}

\section{Introduction}

The properties of the circumnuclear gas in Active Galactic Nuclei (AGNs)
have important consequences on their classification (Antonucci 1993).
Usually AGNs are, optically--wise, divided into type 1 AGNs, showing 
broad permitted emission lines, and type 2 AGNs, which only show narrow 
emission lines.
The Unified Model assumes that AGNs of both classes host the same kind of
nuclear engine and ascribes their differences solely to orientation effects
of a gaseous--dusty torus surrounding the nucleus.
For those lines of sights intercepting the obscuring torus,
both the Broad Line Region (BLR, $< 1$ pc in size) and the nuclear engine
are obscured and only the much more extended
Narrow Line Region (NLR) can be observed. This
model has gained success from a large number of observational tests
(see Antonucci 1993 for a review). In particular, X--ray observations have
supported the unified scenario by discovering large
columns of absorbing gas in type 2 AGNs (e.g. Awaki et al.
1991; Risaliti et al. 1999).
Also, spectroscopic observations in the infrared, where dust absorption is
greatly reduced, detected broad permitted lines in several AGNs which
are classified as type 2 in the optical.

\begin{figure}
\hspace{-0.8cm}
\plottwo{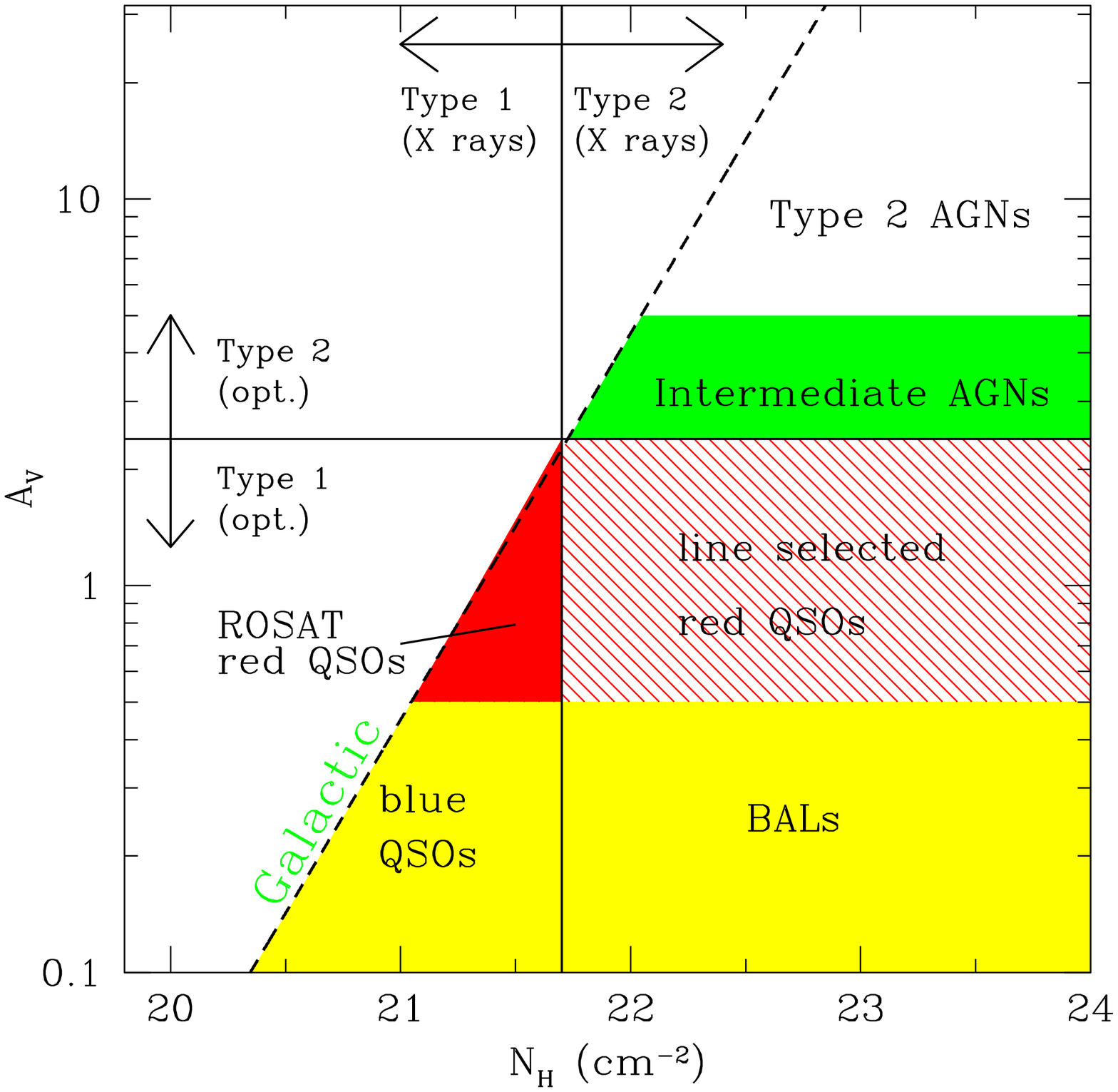}{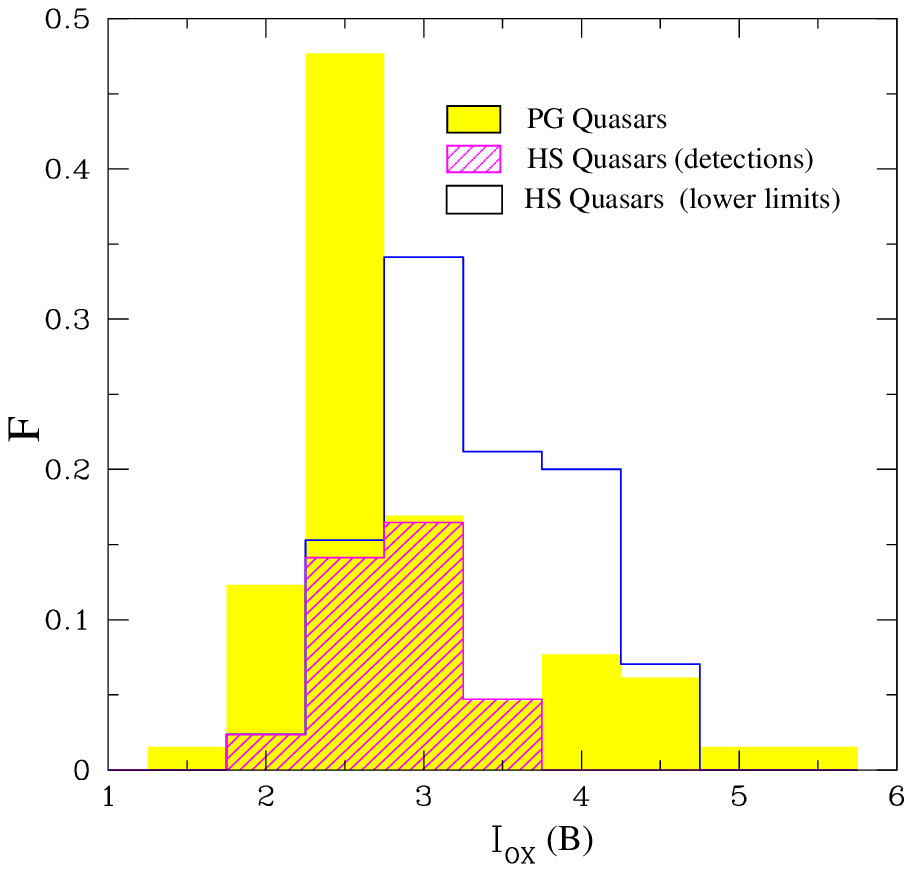}
\caption{$Left$. AGN classification in the $A_{\rm V}-N_{\rm H}$ plane. 
The dashed line represents a Galactic $A_{\rm V}/N_{\rm H}$ value. 
See text for details. $Right$. $I_{\rm OX}$ distribution of the HS QSOs 
compared with that of PG QSOs.}
\end{figure}

However, the properties of the absorbing medium seem to be
more complex, as suggested by the observed mismatch between
the optical and the X--ray classification of some AGNs.
Indeed, most of the blue QSOs showing 
Broad Absorption Lines (BALs) in their UV spectra are underluminous in the 
X--rays, and recently this feature has been ascribed to absorption 
in the X--rays (Brandt et al. 2000).
Also, a significant fraction of hard X--ray selected AGNs with very 
hard X--ray spectra (i.e. probably absorbed) have been subsequently 
identified with type 1 AGNs (Fiore et al. 2000; Akiyama et al. 2000).
Finally, a study of bright QSOs with the ASCA satellite has found evidence 
for X--ray absorption in several of them (Reeves \& Turner 2000).
These findings suggest a mismatch between optical dust absorption and
X--ray gaseous absorption.
In this work we will present evidence for a population of AGNs with 
low $A_{\rm V}/N_{\rm H}$ values. We will also point out that different 
regions of the $A_{\rm V}-N_{\rm H}$ plane are populated by changing the 
selection criteria.

\section{AGN classification in the $A_{\rm V}-N_{\rm H}$ plane}

In Fig.~1 $(left)$ we propose a simple classification scheme for AGNs 
based on 
the absorption effects produced by dust (measured in the optical as 
$A_{\rm V}$) and that produced by metals (both in gas and dust and measured 
in the X--ray band as $N_{\rm H}$). For a dust--to--gas ratio typical of the
diffuse interstellar medium (ISM) of our Galaxy 
$A_{\rm V}(mag)=5\times10^{-22}N_{\rm H}({\rm cm}^{-2})$; this relation 
is represented by the dashed line in Fig.~1. AGNs should be distributed along this line if their circumnuclear gas has the same properties of the Galactic
ISM.
The bottom region of the diagram is populated by objects 
selected by means of their blue colours (i.e. low $A_{\rm V}$) in  
optical surveys (QSOs and BALs). In this diagram BAL QSOs are differentiated 
by ``normal'' blue QSOs due to their large $N_{\rm H}$ inferred 
from the X rays. Selection in the soft X--rays with ROSAT 
has revealed a population of sources with optical colors sufficiently red
to be missed by color based surveys (Kim \& Elvis, 1998).
These objects, showing an $A_{\rm V}/N_{\rm H}$ ratio in agreement 
with the Galactic one, are referred to in Fig.~1 $(left)$ 
as ``ROSAT red QSOs''.
In the next Sections we will present evidence for objects having 
an $A_{\rm V}/N_{\rm H}$ value significantly lower than Galactic 
and populating the regions labelled as ``line selected red QSOs'' and 
``intermediate AGNs'' in Fig.~1 $(left)$.

\section{Line selected red QSOs}

We considered a sample of 
grism--selected AGNs and correlated it with X--ray observations.
Selection on the basis of the optical emission lines should include also 
reddened objects which would have been missed in color based surveys. 
We therefore considered the Hamburg Quasars Sample 
(HS; Engels et al. 1998), whose selection criterion is based on the detection 
of optical/UV broad lines, though a loose color pre--selection is also used.
The HS sample is composed of objects having absolute magnitude $M_{\rm B}<-23$
and therefore any contribution from the host galaxies to the overall luminosity
is negligible.
The HS sample was correlated with the WGACAT X--ray source catalog derived 
from ROSAT PSPC
pointings.\footnote{http://heasarc.gsfc.nasa.gov/W3Browse/all/wgacat.html} 
We found that 85 sources of the HS are in the 
WGACAT fields. Upper limits on the X--ray emission of undetected sources
were calculated according to the exposure time and corrected for
vignetting and PSF.
To avoid the extrapolation to the UV for objects at different redshifts,
we calculated an optical to X--ray index defined as follows, rather than 
using the standard 
$\alpha_{ox}$ index (Avni \& Tananbaum 1986):
\begin{displaymath}
I_{\rm OX}={\rm log}\frac{10^{(20-m_B)/2.5}}{f_x({\rm cts/s})},
\end{displaymath}
where $m_B$ is the B magnitude and $f_x$ is the 0.24--2 keV count rate. 
A detailed description of the sample 
selection and catalogue cross correlation is found in Risaliti et al. (2001).
We compared the $I_{\rm OX}$ distribution of our sample with the same 
distribution 
calculated for the color selected blue QSOs in the Palomar Green sample 
(Schmidt \& Green, 1983). We found that the $I_{\rm OX}$ distribution for the HS
sample and the PG sample are significantly different, the HS distribution being
shifted towards higher values of $I_{\rm OX}$ (i.e. lower X--ray emission relative
to the optical luminosity; see Fig.~1 $right$). We investigated the 
relation of the 
O--E ($\sim$ B--R) optical color
versus redshift for the AGNs in our sample, finding that about 
25\% of the HS QSOs have red O--E colors corresponding to $A_{\rm V}\geq1$
(with a maximum $A_{\rm V}$ of about 2). 
\begin{figure}[t]
\plottwo{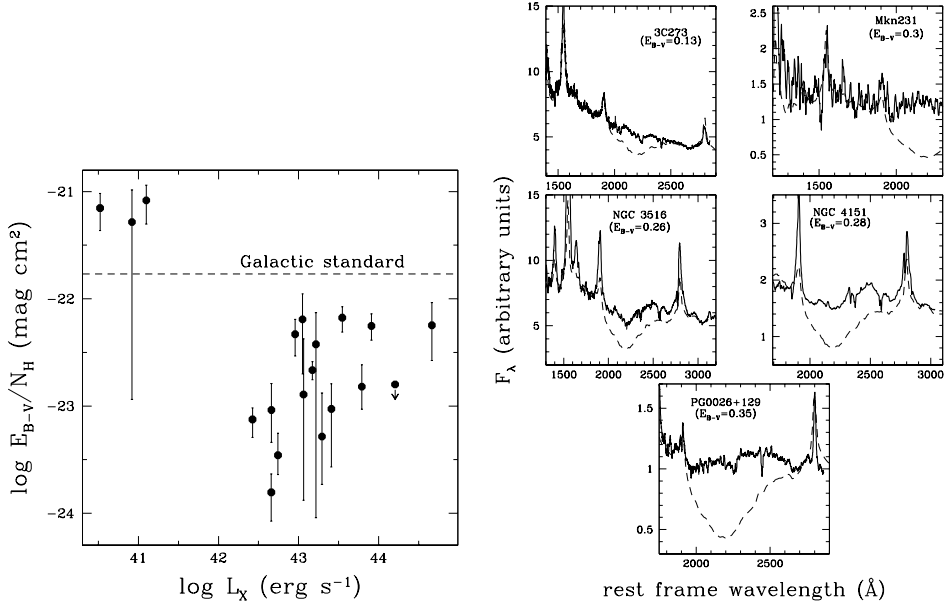}{fig3I.epsi}
\caption{$Left$: $E_{\rm B-V}/N_{\rm H}$ ratio as a function of the intrinsic 
2--10 keV luminosity for the objects in our sample. $Right$: 
UV spectra of five slightly reddened type 1 AGNs. 
The dashed line is the average spectrum of type 1 AGNs
reddened by an $E_{\rm B-V}$ consistent with that measured assuming a standard 
Galactic extinction curve.}
\end{figure}
Therefore, they would have been missed in color based surveys such as the 
PG survey. When comparing the $I_{\rm OX}$ distribution for the HS QSOs with 
$A_{\rm V}\geq1$ (hereafter ``line selected red QSOs'') with that of PG QSOs 
it is found that they are completely different. 
Line selected red QSOs have high $I_{\rm OX}$ indices, suggesting
that absorption is playing a major role in these objects. 
Furthermore, the high values of $I_{\rm OX}$ indicate that absorption is much 
more effective in the X--rays than in the optical, suggesting a
low $A_{\rm V}/N_{\rm H}$ ratio. Therefore, these objects should be
populating the region labelled as ``line selected red QSOs'' 
in Fig.~1 $(left)$.

\section{Intermediate AGNs}

We have defined a sample of AGNs whose X--ray spectrum shows evidence for
cold absorption (hence a measure of the gaseous column density N$_H$
along the line of sight) and whose optical and/or IR spectrum show at
least two {\it broad} lines that are not
completely absorbed by the dust associated to the X--ray absorber.
This sample includes intermediate type (1.8--1.9) Seyferts, type 2 Seyferts
with broad lines in the near--IR, and a few type 1--1.5 Sy characterized
by cold absorption in the X--rays.
The gaseous N$_H$ along the line of sight is derived directly by the 
photoelectric cutoff in the X--ray spectrum.
Ratios between broad components of the hydrogen lines compared
to their intrinsic values give the amount of dust reddening affecting
the BLR. However, radiative transport and collisional excitation effects
in the extreme conditions of the BLR clouds ($\rm n\sim 10^9 cm^{-3}$)
can affect the standard hydrogen line ratios expected in case B
recombination. For instance, BLR models expect the H$\alpha$/H$\beta$ Balmer
decrement to range
from the case B value of 3.1 up to a factor of 3 higher 
(e.g. Mushotzky \& Ferland 1984).
Nonetheless, the H$\alpha$/H$\beta$ ratio observed in Sy1s and QSOs is often
consistent with the standard case B value. Should the
intrinsic ratio be higher, the observed H$\alpha$/H$\beta$ compared to the
case B
value provides at least an upper limit to the reddening.
To determine the reddening E$_{B-V}$ from the broad line ratios we assumed
the ``standard'' Galactic extinction curve (Savage \& Mathis 1979).
As shown in Fig.~2 $(left)$, in ``classical'' AGNs with 
$\rm L_X > 10^{42}\,erg~s^{-1}$ the $E_{\rm B-V}/N_{\rm H}$ ratio is 
significantly 
lower than the Galactic standard value, by a factor ranging from a few to 
$\sim$100. The markedly different behavior of Low Luminosity AGNs 
($\rm L_X < 10^{42}\,erg~s^{-1}$), having $E_{\rm B-V}/N_{\rm H}$ ratios 
higher than
the Galactic value, might reflect intrinsically different physical processes
in these objects with respect to ``classical'' AGNs (e.g. Ho 1999). 
If $A_{\rm V}/E_{\rm B-V}=3.1$ (larger grains have larger ratios, however)
the observed $E_{\rm B-V}$ implies that AGNs with luminosities higher than 
$\rm 10^{42}\,erg~s^{-1}$ 
have $A_{\rm V}=2-4$ and therefore populate the ``intermediate AGNs'' 
region in Fig.~1 $(left)$.

\begin{figure}

\vspace{7cm}
\includegraphics{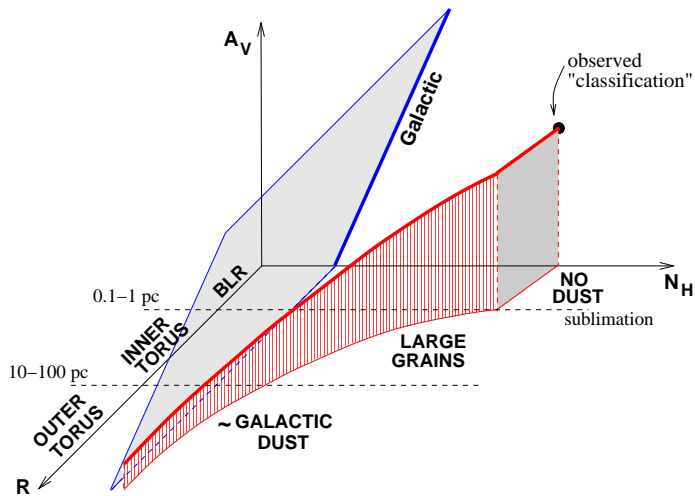}

\caption{A 3D--scheme for the classification of AGNs. This is the analogous
of Fig.~1 ($left$) with an additional dimension represented by the distance
R from the nucleus. The thick curve gives the position on the 
$A_{\rm V}-N_{\rm H}$ plane as a function of R. The thin curve is its
projection on the R--$N_{\rm H}$ plane. The plane labelled as ``Galactic''
represents the locus where $A_{\rm V}/N_{\rm H}$ is Galactic and is shown
for comparison.}
\end{figure}

The low $A_{\rm V}/N_{\rm H}$ values observed in many AGNs could suggest 
dust properties different from those observed in our Galaxy rather than 
a dust--free gas. The large densities of the gas in the vicinity of 
the nucleus could favor the formation of large grains
(Maiolino et al. 2001a). This hypothesis can be tested by looking at 
spectral features produced by small
dust grains such as the absorption dip at 2175\AA \ observed in the diffuse 
interstellar medium and commonly ascribed to small graphite grains with 
radii $\approx 100-200$\AA.
In Fig.~2 $(right)$ we show the UV spectra of five slightly reddened type 1
AGNs for which we could retrieve HST archival spectra.
The thin dashed line shows the template of type 1 AGNs obtained by Francis
et al. (1991) reddened by an $E_{\rm B-V}$ consistent
with that measured from the broad line ratios assuming a 
standard Galactic extinction curve.
The most important result is that the Galactic
extinction curve sistematically predicts a deep feature around 2175\AA \ which
is undetected or much weaker in the observed spectra, suggesting that 
small grains are depleted (Maiolino et al. 2001b).
A dust grain distribution biased in favor of large grains would make the 
extinction curve flatter and featureless (Laor \& Draine 1993). 
If the bias for large 
grains is due to coagulation, this would also explain the observed low 
$E_{\rm B-V}/N_{\rm H}$ and $A_{\rm V}/N_{\rm H}$ ratios (Kim \& Martin 1996).

\section{A structure for absorbers in AGNs}

We suggest that a layered structure of the absorbing medium in AGNs can 
account for the AGN classification in the $A_{\rm V}-N_{\rm H}$ plane.
This is illustrated in Fig.~3, which is the analogous of Fig.~2 $(left)$ with
an additional dimension represented by the distance R from the nucleus.
When R is a few kpc, the density of the absorbing medium
is similar to that of the interstellar medium of the Galaxy, and accordingly
the $A_{\rm V}/N_{\rm H}$ value should be similar to the Galactic one.
As the density of the absorber increases moving towards the nucleus 
(i.e. moving from the outer to the inner torus), the formation of large 
grains is favored and the $A_{\rm V}/N_{\rm H}$ value departs from 
the Galactic value. A further departure could be produced within the Broad
Line Region, where the equilibrium temperature is higher than the sublimation
temperature of the dust and the absorption is therefore provided by 
dust--free gas.
The relative importance of the different absorbing layers should then
be responsible for the distribution of AGNs in the $A_{\rm V}-N_{\rm H}$ plane.

\end{document}